\documentclass[twocolumn,superscriptaddress]{revtex4-1}
\usepackage{graphicx}% Include figure files
\usepackage{dcolumn}% Align table columns on decimal point
\usepackage{multirow}
\usepackage{bm}% bold math
\usepackage{epsfig}% Include figure files
\usepackage{hyperref}

\begin{document}

\title{ Measurement of the $^{58}$Ni($\alpha$,$\gamma$)$^{62}$Zn reaction and its astrophysical impact }

\author{S.~J.~Quinn}
   \email[]{quinn@nscl.msu.edu}
   \affiliation{National Superconducting Cyclotron Laboratory, Michigan State University, East Lansing, MI 48824, USA}
   \affiliation{Department of Physics \& Astronomy, Michigan State University, East Lansing, MI 48824, USA}
   \affiliation{Joint Institute for Nuclear Astrophysics, Michigan State University, East Lansing, MI 48824, USA}
\author{A.~Spyrou}
   \affiliation{National Superconducting Cyclotron Laboratory, Michigan State University, East Lansing, MI 48824, USA}
   \affiliation{Department of Physics \& Astronomy, Michigan State University, East Lansing, MI 48824, USA}
   \affiliation{Joint Institute for Nuclear Astrophysics, Michigan State University, East Lansing, MI 48824, USA}
\author{E.~Bravo}
   \affiliation{Dept. F\'isica i Enginyeria Nuclear, Univ. Polit\`ecnica de Catalunya, Carrer Pere Serra 1-15, 08173 Sant Cugat del Vall\`es, Spain}
\author{T.~Rauscher}
	\affiliation{Centre for Astrophysics Research, School of Physics, Astronomy, and Mathematics, University of Hertfordshire, Hatfield AL10 9AB, United Kingdom}
	\affiliation{Department of Physics, University of Basel, 4056 Basel, Switzerland}
\author{A.~Simon}
	 %\altaffiliation{Department of Physics, University of Richmond, Richmond, Virginia 23173, USA}
	 \altaffiliation{current address: Department of Physics, University of Richmond, Richmond, VA 23173, USA}
   \affiliation{National Superconducting Cyclotron Laboratory, Michigan State University, East Lansing, MI 48824, USA}
   \affiliation{Joint Institute for Nuclear Astrophysics, Michigan State University, East Lansing, MI 48824, USA}

\author{A.~Battaglia}
   \affiliation{Department of Physics and The Joint Institute for Nuclear Astrophysics, University of Notre Dame, Notre Dame, IN 46556, USA}
\author{M.~Bowers}
   \affiliation{Department of Physics and The Joint Institute for Nuclear Astrophysics, University of Notre Dame, Notre Dame, IN 46556, USA}
\author{B.~Bucher}
   \affiliation{Department of Physics and The Joint Institute for Nuclear Astrophysics, University of Notre Dame, Notre Dame, IN 46556, USA}
\author{C.~Casarella}
   \affiliation{Department of Physics and The Joint Institute for Nuclear Astrophysics, University of Notre Dame, Notre Dame, IN 46556, USA}
\author{M.~Couder}
   \affiliation{Department of Physics and The Joint Institute for Nuclear Astrophysics, University of Notre Dame, Notre Dame, IN 46556, USA}
\author{P.~A.~DeYoung}
   \affiliation{Department of Physics, Hope College, Holland, MI 49423, USA}
\author{A.~C.~Dombos}
   \affiliation{National Superconducting Cyclotron Laboratory, Michigan State University, East Lansing, MI 48824, USA}
   \affiliation{Department of Physics \& Astronomy, Michigan State University, East Lansing, MI 48824, USA}
   \affiliation{Joint Institute for Nuclear Astrophysics, Michigan State University, East Lansing, MI 48824, USA}
\author{J.~G\"{o}rres}
   \affiliation{Department of Physics and The Joint Institute for Nuclear Astrophysics, University of Notre Dame, Notre Dame, IN 46556, USA}
\author{A.~Kontos}
   \affiliation{National Superconducting Cyclotron Laboratory, Michigan State University, East Lansing, MI 48824, USA}
   \affiliation{Joint Institute for Nuclear Astrophysics, Michigan State University, East Lansing, MI 48824, USA}
   \affiliation{Department of Physics and The Joint Institute for Nuclear Astrophysics, University of Notre Dame, Notre Dame, IN 46556, USA}
\author{Q.~Li}
   \affiliation{Department of Physics and The Joint Institute for Nuclear Astrophysics, University of Notre Dame, Notre Dame, IN 46556, USA}
\author{A.~Long}
   \affiliation{Department of Physics and The Joint Institute for Nuclear Astrophysics, University of Notre Dame, Notre Dame, IN 46556, USA}
\author{M.~Moran}
   \affiliation{Department of Physics and The Joint Institute for Nuclear Astrophysics, University of Notre Dame, Notre Dame, IN 46556, USA}
\author{N.~Paul}
   \affiliation{Department of Physics and The Joint Institute for Nuclear Astrophysics, University of Notre Dame, Notre Dame, IN 46556, USA}
\author{J.~Pereira}
   \affiliation{National Superconducting Cyclotron Laboratory, Michigan State University, East Lansing, MI 48824, USA}
\author{D.~Robertson}
   \affiliation{Department of Physics and The Joint Institute for Nuclear Astrophysics, University of Notre Dame, Notre Dame, IN 46556, USA}
\author{K.~Smith}
   \affiliation{Department of Physics and The Joint Institute for Nuclear Astrophysics, University of Notre Dame, Notre Dame, IN 46556, USA}
\author{M.~K.~Smith}
   \affiliation{Department of Physics and The Joint Institute for Nuclear Astrophysics, University of Notre Dame, Notre Dame, IN 46556, USA}
\author{E.~Stech}
   \affiliation{Department of Physics and The Joint Institute for Nuclear Astrophysics, University of Notre Dame, Notre Dame, IN 46556, USA}
\author{R.~Talwar}
   \affiliation{Department of Physics and The Joint Institute for Nuclear Astrophysics, University of Notre Dame, Notre Dame, IN 46556, USA}
\author{W.~P.~Tan}
   \affiliation{Department of Physics and The Joint Institute for Nuclear Astrophysics, University of Notre Dame, Notre Dame, IN 46556, USA}
\author{M.~Wiescher}
   \affiliation{Department of Physics and The Joint Institute for Nuclear Astrophysics, University of Notre Dame, Notre Dame, IN 46556, USA}

%\date{\today}

\begin{abstract}

Cross section measurements of the $^{58}$Ni($\alpha$,$\gamma$)$^{62}$Zn reaction were performed in the energy range
$E_{\alpha}=5.5-9.5$ MeV at the Nuclear Science Laboratory of the University of Notre Dame, using the NSCL Summing NaI(Tl) detector and the
$\gamma$-summing technique. The measurements are compared to predictions in the statistical Hauser-Feshbach model of nuclear reactions using the
SMARAGD code. It is found that the energy dependence of the cross section is reproduced well but the absolute value is overestimated by the prediction. This can be remedied by rescaling the $\alpha$ width by a factor of 0.45. Stellar reactivities were calculated with the rescaled $\alpha$ width and their impact on nucleosynthesis in type Ia supernovae has been studied. It is found that the resulting abundances change by up to 5\% when using the new reactivities.

% \begin{description}
%   \item[PACS numbers] {24.60.Dr, 25.40.Lw, 26.30.-k, 27.40.+z}
% \end{description}

\end{abstract}

\maketitle

%\linenumbers
%\setlength\linenumbersep{0.15cm}

%---------------------------------------------------------------------------------------------------------------------------------------------
\section{Introduction}
\label{sec:intro}

For a full understanding of nulceosynthesis, astronomical observations and stellar modeling  must be combined with  
nuclear physics measurements (e.g.~\cite{Kappeler11, Arnould07, Rauscher13}).  By measuring reaction cross sections at astrophysical energies,
reaction rates can be determined and used in nucleosynthetic codes to predict isotopic abundances in various stellar environments.
However, in many cases experimental data do not exist and instead theoretical rates and their uncertainties are relied upon.
It is therefore important to provide experimental data when possible, either to be used directly in astrophysical calculations or
for constraining theoretical models and improving their predictive power.

To understand the impact of a specific reaction in a nucleosynthesis process, the reaction rate can be varied within the
uncertainties of the theory while tracking how the final isotopic abundances are altered
(see, e.g.,~\cite{The98, Iliadis02, Hix03, Rapp06}). One such sensitivity study was recently performed by Bravo and
Mart\'inez-Pinedo~\cite{Bravo12} to quantitatively understand the influence of individual reaction rates on the
nucleosynthesis in type Ia supernovae (SNIa).  For the study, the authors used a one-dimensional delayed detonation model
of a Chandrasekhar-mass white dwarf and varied the reaction rates by a factor of 10 up and down. Overall, the authors
concluded that nucleosynthesis was relatively insensitive to the change of individual reaction rates, but many reactions
were identified as relevant for having a direct impact on the final abundance of particular isotopes. One such reaction
was the $^{58}$Ni($\alpha$,$\gamma$)$^{62}$Zn reaction, which was selected for its impact on the production of
$^{62}$Ni, $^{63}$Cu, and $^{64}$Zn.

The astrophysical scenario in which the $^{58}$Ni($\alpha$,$\gamma$)$^{62}$Zn reaction is expected to play an important role
is during the $\alpha$-rich freeze-out from nuclear statistical equilibrium (NSE).  In the innermost layers of SNIa, temperatures
and densities are sufficiently high to reach NSE, and a large portion of the material in NSE is expected to undergo $\alpha$-rich freeze-out.
For the Chandrasekhar-mass white dwarf studied in Ref.~\cite{Bravo12} it was noticed that the inner 0.4M$_{\odot}$ reached NSE,
of which 0.24M$_{\odot}$ underwent $\alpha$-rich freeze-out.  After $\alpha$-rich freeze-out, the composition of the layer is
dominated by isotopes in the iron region and $\alpha$-particles that did not reassemble into heavier nuclei.  Thus it is
expected that $\alpha$-induced reactions on nuclei in the iron region are important for the final abundance pattern.

The $^{58}$Ni($\alpha$,$\gamma$)$^{62}$Zn reaction has been measured three times previously all over 50 years ago.
Morinaga~\cite{Morinaga56} and Ball et.al.~\cite{Ball59} performed cross section measurements using the activation
technique with energies E$_{\alpha}$~=~10.6~-~31.0~MeV. After irradiation, both measurements included an additional step
of chemically seperating zinc from other elements before counting the decay of $^{62}$Zn with Geiger counters.  McGowan and
collaborators~\cite{McGowan64} extended the measurements to lower energies by using thick-target yields
from enriched $^{58}$Ni targets. The yield was determined every 100~keV within the beam energy range of E$_{\alpha}$~=~4.9~-~6.1~MeV
and the cross section determined by differentiating the yield curve. In the present work, we report on a
new measurement of the $^{58}$Ni($\alpha$,$\gamma$)$^{62}$Zn reaction cross section using the $\gamma$-summing
technique. The new values serve to verify the
previous results as well as to expand the energy coverage of experimental measurements.  The larger energy coverage gives an improved
understanding of the energy dependence of the cross section  and allows for better comparison to theoretical models.
In Sec.~\ref{sec:experiment} of this paper we describe the experimental setup and we provide the experimental results in
Sec.~\ref{sec:results}.  In Sec.~\ref{sec:discussion} we compare the measurements to theoretical predictions and provide new
reactivities for the $^{58}$Ni($\alpha$,$\gamma$)$^{62}$Zn reaction. Lastly, in Sec.~\ref{sec:astro} we investigate the application of the new
reaction rates to SNIa nucleosynthesis.

%---------------------------------------------------------------------------------------------------------------------------------------------
\section{Experimental details}
\label{sec:experiment}

The experiment utilized the FN Tandem Van de Graaff Accelerator at the University of Notre Dame to accelerate $^{4}$He$^{2+}$
nuclei to energies $E_{\alpha}=5.5-9.5$ MeV. The beam current was varied between
$4-60$ enA in order to balance count rate with minimal detection dead time. For the present work the dead time was kept below 1.2\%.
The total charge collected was between $7-159$ $\mu$C
for each data run as determined by a Faraday cup at the end of the beamline.

The $^{58}$Ni target was isotopically enriched to 95(5)\% and its thickness was measured using Rutherford Backscattering 
Spectrometry (RBS) performed at the Hope College
Ion Beam Analysis Laboratory (HIBAL)~\cite{HIBAL}. The experimental setup for the RBS measurements consisted of a silicon surface
barrier detector with a 0.2 inch diameter collimator placed at 168.2 degrees in respect to the incoming 2.94~MeV $^{4}$He$^{2+}$ beam.
The resulting backscattered spectra were fit with SIMNRA software~\cite{Mayer97} with the $^{58}$Ni target composition
and thickness as free parameters. The thickness was determined to be $930 \pm 46$ $\mu$g/cm$^{2}$. Thus the energy loss was
 0.42~MeV and 0.30~MeV at $E_{\alpha}=5.5$ MeV and $E_{\alpha}=9.5$ MeV, respectively~\cite{LISE}.  Trace amounts of carbon and oxygen 
were seen on the front and back surfaces of the target.

To perform the cross section measurements the $^{58}$Ni target was mounted at the center
of the Summing NaI(Tl) (SuN) detector from the National Superconducting Cylotron Laboratory (NSCL) of Michigan State
University~\cite{Simon13}.  The cylindrical SuN detector is 16 inches in diameter and 16 inches in length with a 1.77 inch diameter
borehole along its axis. The entire volume of SuN is divided into 8 semi-cylindrical segments which are optically isolated and
read out by 3 photomultiplier tubes each. The signals are recorded using the NSCL Digital Data Acquistion System
(DDAS)~\cite{Prokop14}. The nearly 4$\pi$ angular coverage provided by SuN allows cross sections to be measured via the
$\gamma$-summing technique~\cite{Simon13, Spyrou07}. In this technique, the deexcitation $\gamma$ rays from the produced
nuclei are detected and summed up to an energy equal to the entry state. Thus, instead of analyzing individual $\gamma$-ray
transitions only the ``sum peak'' corresponding to the sum of the sequential $\gamma$ rays needs to be integrated. The sum peak
is located at an energy of $E_{\Sigma}=E_\mathrm{c.m.}+Q$ in the $\gamma$-ray spectrum,
where $E_\mathrm{c.m.}$ is the center of mass energy of the projectile-target system and $Q$ is the reaction $Q$ value.

\begin{figure}
\begin{center}
\includegraphics[width=\columnwidth]{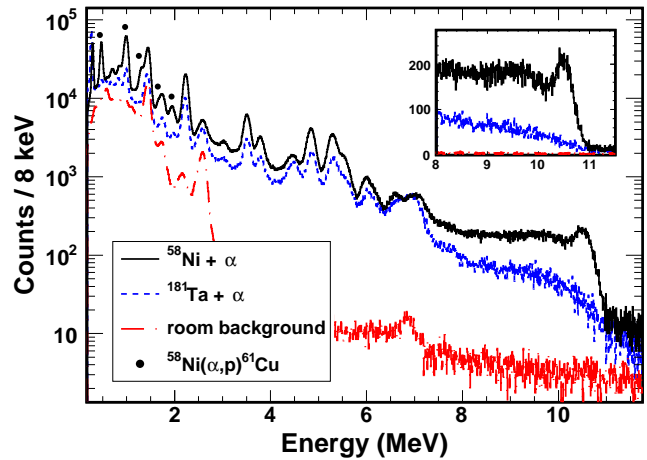}
\end{center}
\caption{\textit{(color online)} Experimental spectra from the SuN detector for measurements at $E_{\alpha}=7.7$ MeV.
The spectra correspond to $^{58}$Ni (solid black), thick tantalum backing (dotted blue), and normalized room background
(dashed red). The inset shows a zoom around the sum-peak region of the $^{58}$Ni($\alpha$,$\gamma$)$^{62}$Zn reaction.}
\label{fig:sumPeak}
\end{figure}

A spectrum from the $^{58}$Ni($\alpha$,$\gamma$)$^{62}$Zn reaction with $E_\mathrm{c.m.}=7039$ keV and $Q=3364.27$ keV is
shown in Fig. \ref{fig:sumPeak}.  At higher energies, both room background and beam-induced background contributions
to the spectrum are reduced allowing the sum peak to be clearly visible at 10.4~MeV. The source of room background in
the region of the sum peak comes from cosmic rays. During the experiment the $^{58}$Ni target was mounted in front
of a thick tantalum backing.  Thus, the beam induced background was determined by taking data without the
$^{58}$Ni target in place so that the beam was impinging solely onto the tantalum backing. Additional peaks were
visible in the low energy region of the $^{58}$Ni spectrum that originate from the $^{58}$Ni($\alpha$,p$\gamma$)$^{61}$Cu reaction
which has a higher cross section than the $^{58}$Ni($\alpha$,$\gamma$)$^{62}$Zn reaction at this energy by apporoximately
two orders of magnitude~\cite{McGowan64,Vlieks74}. The additional nickel isotopes have ($\alpha$,$\gamma$) $Q$ values larger than the 
$^{58}$Ni($\alpha$,$\gamma$)$^{62}$Zn reaction, and thus do not contribute in the sum-peak region. Also, these additional nickel 
isotopes are present in very low amounts in the target and there was no indication of their ($\alpha$,$\gamma$) reactions in the 
summed spectra.

\begin{table}
\caption{Cross sections %and astrophysical S-factors
for the $^{58}$Ni($\alpha$,$\gamma$)$^{62}$Zn reaction.}
\begin{ruledtabular}
\begin{tabular}{c c dD{,}{\pm}{-1}}
%\begin{tabular}{c c c c}
\multicolumn{1}{c}{$E_\mathrm{c.m.}^\mathrm{max}$ (MeV)} & 
\multicolumn{1}{c}{$E_\mathrm{c.m.}^\mathrm{min}$ (MeV)} &
\multicolumn{1}{c}{$E_\mathrm{c.m.}^\mathrm{eff}$ (MeV)} &
\multicolumn{1}{c}{$\sigma$ ($\mu$b)} \\
\hline
5.143	&	4.750	&	4.988	&	3.13	,	0.44 \\
5.330	&	4.946	&	5.171	&	4.70	,	0.60 \\
5.517	&	5.143	&	5.360	&	6.69	,	1.04 \\
5.704	&	5.337	&	5.548	&	9.65	,	1.34 \\
6.078	&	5.723	&	5.922	&	15.3	,	2.4 \\
6.452	&	6.112	&	6.298	&	22.2	,	3.5 \\
6.826	&	6.496	&	6.673	&	34.0	,	6.1 \\
7.201	&	6.883	&	7.051	&	52.4	,	7.1 \\
7.574	&	7.268	&	7.428	&	66.9	, 9.7	\\
7.949	&	7.649	&	7.805	&	92.8	,	15.1\\
8.415	&	8.129	&	8.277	&	138.8	,	22.1\\
8.884	&	8.606	&	8.749	&	158.9	,	25.2
\end{tabular}
\end{ruledtabular}
\label{table:data}
%\end{center}
\end{table}

%---------------------------------------------------------------------------------------------------------------------------------------------
\section{Results}
\label{sec:results}

The $^{58}$Ni($\alpha$,$\gamma$)$^{62}$Zn cross section was calculated from
\begin{equation}
\label{eq:xsection}
\sigma = \frac{ N_{\Sigma} }{ N_{\alpha}  n_{t}  \varepsilon_{\Sigma} }
\end{equation}
where $N_{\Sigma}$ is the number of counts in the sum peak, $N_{\alpha}$ is the number of projectiles,
$n_\mathrm{t}$ is the areal target density, and $\varepsilon_{\Sigma}$ is the sum-peak efficiency.
In the present work, $N_{\alpha}$ was measured with a Faraday cup and current integrator and $n_\mathrm{t}$ was
measured with the RBS technique as mentioned in Sec.~\ref{sec:experiment}.  To determine $N_{\Sigma}$,
a linear background was subtracted and the sum peak was integrated in the region of 3 standard deviations
below and above the sum peak centroid. This method was chosen to be consistent with the efficiency
calculations of Ref.~\cite{Simon13}.  Finally, $\varepsilon_{\Sigma}$ depends not only on the energy
but also on the average number of $\gamma$ rays emitted, or ``multiplicity'', of the cascade from the
entry state to the final state. The multiplicity is not known beforehand but can be experimentally
determined by calculating the average number of segments in SuN that detect $\gamma$-ray energy for a
sum peak event. The number of segments participating in a sum peak event is referred to as the
``hit pattern'' and more details on the technique can be found in Ref.~\cite{Simon13}.  The efficiencies
were determined to range from 26.7(2.8)\% at $E_\mathrm{c.m.}=4.943$ MeV to 17.4(2.3)\% at $E_\mathrm{c.m.}=8.742$ MeV.

The results of the present work are displayed in Table~\ref{table:data}. In the table, the first two columns contain the
maximum and minimum energies of the beam due to the thickness of the target. The third column contains the effective energy 
for each data point taking into account the variation of the cross section in the target. The last column lists the cross 
sections calculated from Eq.~(\ref{eq:xsection}). 
Of the uncertainty reported, roughly 3\% comes from statistical uncertainties, 5\% from the beam
charge collection, 5\% from the target thickness, 5\% from the target enrichment, and 10\%~-~15\% from the detection 
efficiency. The uncertainty in energy from the accelerator is 4 keV at all energies.

\begin{figure}
\begin{center}
\includegraphics[width=\columnwidth]{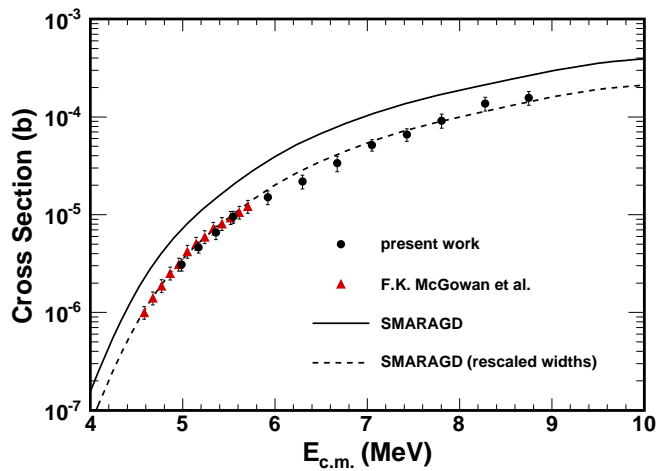}
\end{center}
\caption{\textit{(color online)} Cross section of the $^{58}$Ni($\alpha$,$\gamma$)$^{62}$Zn reaction for the present
work (black circles), previous data of Ref.~\cite{McGowan64} (red triangles), and theoretical calculations from the SMARAGD
code~\cite{SMARAGD}. A good description was obtained by modifying the $\alpha$ width and the $\gamma$-to-proton width ratio (dashed line).}
\label{fig:xsection}
\end{figure}

A plot of the $^{58}$Ni($\alpha$,$\gamma$)$^{62}$Zn reaction cross section %for astrophysically relevant energies
is shown in Fig.~\ref{fig:xsection}. The present work is in agreement with the previous results of
Ref.~\cite{McGowan64} and extends the measurements to higher energies. 
The increased energy coverage of the experimental cross section allows for a more sensitive study of the energy dependence of the 
cross section and provides a better test for theoretical models described in the next section

\begin{figure}
\begin{center}
\includegraphics[width=\columnwidth]{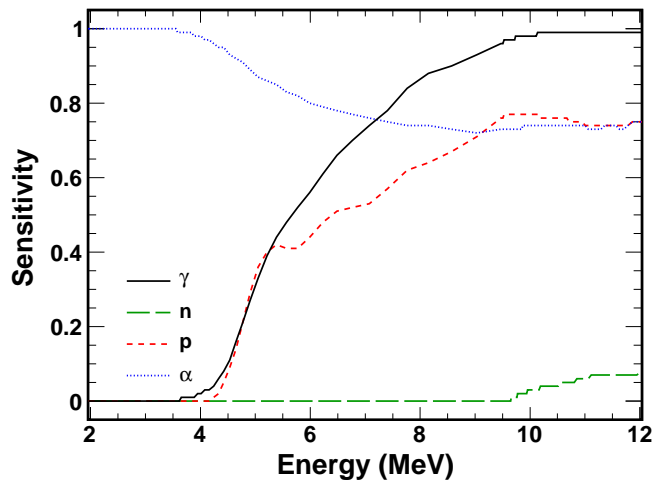}
\end{center}
\caption{\textit{(color online)} Absolute values of the sensitivity of the $^{58}$Ni($\alpha$,$\gamma$)$^{62}$Zn cross section as function of energy, when separately varying $\gamma$, neutron, proton, and $\alpha$~widths.}
\label{fig:sensitivity}
\end{figure}

%---------------------------------------------------------------------------------------------------------------------------------------------
\section{Discussion}
\label{sec:discussion}

%The theoretical investigation of the $^{58}$Ni($\alpha$,$\gamma$)$^{62}$Zn reaction begins by understanding the
%sensitivity of the cross section to several key input parameters in the nuclear statistical model. 

The theoretical investigation of the $^{58}$Ni($\alpha$,$\gamma$)$^{62}$Zn reaction was performed using the nuclear statistical model. 
The limit of applicability of the statistical model in this case is 0.12 GK~\cite{prc56}, which is well below the relevant 
temperature range of 2-5 GK for the astrophysical applications. The corresponding Gamow window at 2 GK is from approximately 
3 to 5 MeV with a maximum contribution to the rate 
at 4 MeV, and at 5 GK the Gamow window is from 4 to 7 MeV with a maximum contribution at 5.25 MeV~\cite{energywindows}. 
While the experimental values cover the upper part of the energy window, theoretical predictions are required at lower energies to better 
constrain the reaction rate.

In Fig.~\ref{fig:sensitivity},  the sensitivity of the 
$^{58}$Ni($\alpha$,$\gamma$)$^{62}$Zn cross section to variations in the $\gamma$, neutron, proton, and 
$\alpha$ widths, respectively, is shown. The relative sensitivity is defined as \cite{Rauscher12}
\begin{equation}
\label{eq:sensi}
\Omega_{S_q} = \frac{\upsilon_\Omega - 1}{\upsilon_q - 1}
\end{equation}
where $q$ is the quantity being changed and $\Omega$ is the resulting quantity.
A change in $q$ is given by the factor $\upsilon_q=q_{new}/q_{old}$ and the subsequent change in $\Omega$ is given
by the factor $\upsilon_\Omega=\Omega_{new}/\Omega_{old}$. Using these definitions, it is clear that the sensitivity
$\Omega_{S_q}=0$ when no change occurs and $\Omega_{S_q}=1$ when $\Omega$ changes by the same factor as $q$. In the current context, the quantity $q$ is an averaged width as used in the reaction model (Sec.\ \ref{sec:discussion}) or a stellar reactivity as used in the nucleosynthesis model (Sec.\ \ref{sec:astro}). Then the resulting quantity $\Omega$ is the cross section and the abundance of an isotope, respectively.

Below 4~MeV the $^{58}$Ni($\alpha$,$\gamma$)$^{62}$Zn reaction cross section is exclusively sensitive to the $\alpha$
width with the sensitivity to the $\alpha$ width persisting throughout the energy region plotted. 
This low-energy region is also important for the calculation of the astrophysical reaction rate and reactivity \cite{energywindows}. 
Conversely, there is very little sensitivity to the neutron width even for energies above the neutron emission threshold at $E_\mathrm{c.m.}=9.526$ MeV. 
The remaining two parameters, the proton and $\gamma$ widths, show an increasing effect on the cross section with increasing energy in the region between 
4 and 10~MeV.
% The temperature range for the astrophysical application is 2 - 5 GK.  
 
For the present work, theoretical calculations were performed with the Hauser-Feshbach reaction model \cite{haufesh,raureview} using the code SMARAGD~\cite{SMARAGD,raureview}. The initial calculation systematically overestimated the cross section values, as shown with the solid line of Fig.~\ref{fig:xsection}, although the energy dependence was reproduced well. Adapting the particle and $\gamma$ widths, good agreement with the data across the measured energy range was achieved. As can be seen from the sensitivities in Fig.\ \ref{fig:sensitivity} the $\alpha$ width is constrained by the low-energy data, below 6 MeV, requiring a rescaling of the width obtained with the optical potential of \cite{McFadden66} by a factor 0.45. The $\gamma$- and proton widths cannot be constrained separately with data from only this reaction. Increasing the $\gamma$-to-proton width ratio by 10\% provides improved agreement with the data at the upper end of the measured energy range, above 7 MeV. The calculation with the rescaled widths is shown as dashed line in Fig.~\ref{fig:xsection}). 
Although the scaling factors for the $\alpha$- width and $\gamma$-to-proton width ratio provide an excellent description of the $^{58}$Ni($\alpha$,$\gamma$)$^{62}$Zn cross section, calculations using these scaling factors underproduce the $^{58}$Ni($\alpha$,p)$^{61}$Cu experimental data \cite{McGowan64,Vlieks74} by a factor of three.  Further theoretical work is required to obtain a full understanding of $\alpha$-induced reaction cross sections on $^{58}$Ni.

Since there is no indication from our data that the energy dependence changes towards even lower energies and the data by \cite{McGowan64} are 
also reproduced well, we used the modified widths to predict cross sections at such lower, unmeasured energies and calculated new stellar 
reactivities, which are shown in Table~\ref{table:rates}. These stellar reactivities are dominated by the ground-state cross sections with only 
small influence from thermally excited states in $^{58}$Ni.  The ground-state contributions are 100\% at 2 GK and 95\% at 5 GK and thus the 
reactivities are well constrained by experimental data.  In Table \ref{table:parameters} we also provide a fit to the stellar reactivities in 
the standard 7 parameter REACLIB format \cite{Rauscher00} as commonly used in astrophysical calculations.

\begin{table}
\caption{Stellar reactivities for the $^{58}$Ni($\alpha$,$\gamma$)$^{62}$Zn reaction.}
\begin{center}
\begin{tabular}{c @{\hskip 0.2in} c @{\hskip 0.4in} c @{\hskip 0.2in} c}
\hline\hline
T		& Reactivity 							& T 		& 	Reactivity 							\\
(GK) 	& (cm$^{3}$ mol$^{-1}$ s$^{-1}$)	& (GK) 	&  (cm$^{3}$ mol$^{-1}$ s$^{-1}$)\\ [0.5ex]
\hline
0.10	&	6.623	$\times$ 10$^{-62}$		&	2.00	&	2.268	$\times$ 10$^{-7}$		\\
0.15	&	1.971	$\times$ 10$^{-50}$		&	2.50	&	2.045	$\times$ 10$^{-5}$		\\
0.20	&	2.068	$\times$ 10$^{-43}$		&	3.00	&	5.000	$\times$ 10$^{-4}$		\\
0.30	&	1.171	$\times$ 10$^{-34}$		&	3.50	&	5.407	$\times$ 10$^{-3}$		\\
0.40	&	3.546	$\times$ 10$^{-29}$		&	4.00	&	3.399	$\times$ 10$^{-2}$		\\
0.50	&	2.649	$\times$ 10$^{-25}$		&	4.50	&	1.461	$\times$ 10$^{-1}$		\\
0.60	&	2.293	$\times$ 10$^{-22}$		&	5.00	&	4.755	$\times$ 10$^{-1}$		\\
0.70	&	4.941	$\times$ 10$^{-20}$		&	6.00	&	2.812	$\times$ 10$^{0}$			\\
0.80	&	4.070	$\times$ 10$^{-18}$		&	7.00	&	9.844	$\times$ 10$^{0}$			\\
0.90	&	1.666	$\times$ 10$^{-16}$		&	8.00	&	2.443	$\times$ 10$^{1}$			\\
1.00	&	4.020	$\times$ 10$^{-15}$		&	9.00	&	4.755	$\times$ 10$^{1}$			\\
1.50	&	2.679	$\times$ 10$^{-10}$		&	10.00	&	7.688	$\times$ 10$^{1}$			\\ [0.5ex]
\hline
\end{tabular}
\label{table:rates}
\end{center}
\end{table}

\begin{table}
\caption{REACLIB parameters for the $^{58}$Ni($\alpha$,$\gamma$)$^{62}$Zn reaction. }
\begin{center}
\begin{tabular}{c r l}
\hline\hline
\multicolumn{1}{c}{Parameter} & \multicolumn{2}{c}{Value} \\ [0.5ex]
\hline
$a_0$	&	 	&	5.194217 $\times$ 10$^{1}$	 \\
$a_1$	&	$-$&	2.314329	$\times$ 10$^{0}$	 \\
$a_2$	&	$-$& 2.528868 $\times$ 10$^{1}$	 \\
$a_3$	&	$-$& 5.651307 $\times$ 10$^{1}$	 \\
$a_4$	&	$-$& 1.088296	$\times$ 10$^{0}$	 \\
$a_5$	&	 	& 1.763373 $\times$ 10$^{-1}$ \\
$a_6$	&	 	&	3.858753 $\times$ 10$^{1}$	 \\ [0.5ex]
\hline
\end{tabular}
\label{table:parameters}
\end{center}
\end{table}

%---------------------------------------------------------------------------------------------------------------------------------------------
\section{Astrophysical Calculations}
\label{sec:astro}

%EBstart
The effect of the new reactivities on the nucleosynthesis of type Ia supernovae was also
investigated. A reduction in the rate of $^{58}$Ni($\alpha$,$\gamma$)$^{62}$Zn is expected to
translate into a decrease in the abundance of $^{62}$Zn and other nuclei linked by subsequent
reaction chains, e.g. $^{62}$Zn($\alpha$,$\gamma$)$^{66}$Ge,
$^{62}$Zn(p,$\gamma$)$^{63}$Ga(p,$\gamma$)$^{64}$Ge, and so on. After disintegration of the
radioactive isotopes, the result is a decrease of the ejected abundances of $^{62}$Ni, $^{66}$Zn,
$^{63}$Cu, $^{64}$Zn, and others.
In \cite{Bravo12}, it was determined that the relative sensitivity of the ejected
mass of $^{62}$Ni with respect to the $^{58}$Ni($\alpha$,$\gamma$)$^{62}$Zn rate is $\Omega_{S_q} =
0.12$. In the temperature range in which $\alpha$-rich freeze-out takes place, namely from
$\sim2$~GK to $\sim5$~GK, the new reaction rates given by Table~\ref{table:parameters} are smaller
than the rates compiled in \cite{Cyburt10} by a factor of $\upsilon_q~\sim~0.45$.
Thus,
the expected change in the abundance of $^{62}$Ni related to the new rates is $\sim6-7$\%.

We have conducted simulations of type Ia supernovae with both the rates of Ref.~\cite{Cyburt10} and the
new rates following the same methodology and codes as described in \cite{Bravo12}.
Table~\ref{table:SNIaresults} shows the relative changes in the ejected abundances of the most
sensitive species for both a delayed detonation of a Chandrasekhar-mass white dwarf and a
thermonuclear explosion of a sub-Chandrasekhar white dwarf. The results agree with the prior
estimate with a maximum sensitivity of $\sim5$\% for the abundances of $^{62}$Ni and $^{64}$Zn.
We have repeated the calculations with different sets of deflagration-to-detonation transition
densities, $\rho_\mathrm{DDT}$, and initial metallicities, but their effect is small and the maximum
sensitivities never exceed the values reported in Table~\ref{table:SNIaresults} (in general, the
sensitivities increase with metallicity and with $\rho_\mathrm{DDT}$).

\begin{table}
\caption{Changes to the nucleosynthesis of type Ia supernova models.\label{table:SNIaresults}}
%\begin{center}
\begin{ruledtabular}
\begin{tabular}{ccc}
%\hline\hline
$\upsilon_\Omega -1$ & Delayed detonation\footnotemark[1] & SubChandrasekhar\footnotemark[2] \\
%[0.5ex]
\hline
-0.05 & $^{64}$Zn & $^{62}$Ni \\
-0.04 & $^{62}$Ni &  \\
-0.02 & $^{63}$Cu, $^{66}$Zn & $^{63}$Cu, $^{66}$Zn, $^{69}$Ga \\
-0.01 & & $^{64}$Zn, $^{65}$Zn, $^{73}$Ge %\\ [0.5ex]
%\hline
\end{tabular}
\end{ruledtabular}
\footnotetext[1]{Chandrasekhar-mass delayed detonation model with $\rho_\mathrm{DDT} =
3.9\times10^7$~g/cm$^3$.}
\footnotetext[2]{Explosion of a sub-Chandrasekhar white dwarf of 1.025~M$_\odot$ C-O core
surrounded by a 0.055~M$_\odot$ He envelope.}
%\end{center}
\end{table}
%EBend

%---------------------------------------------------------------------------------------------------------------------------------------------
\section{Conclusions}
\label{sec:conclusion}

The $^{58}$Ni($\alpha$,$\gamma$)$^{62}$Zn reaction cross section was measured for the particle energies from $E_{\alpha}=5.5$ MeV
to $E_{\alpha}=9.5$ MeV at the University of Notre Dame. The measurements were performed using the SuN
detector and $\gamma$-summing technique. The present results agree well with previous measurements and the
data was compared to theoretical calculations using the nuclear statistical model.  The standard calculation
by the SMARAGD code overproduces the measured cross section, but multiplying the $\alpha$ width by a factor
of 0.45 accurately reproduces the data.  New reactivities were reported, and the new reaction rates used
in nucleosynthesis calculations for type Ia supernovae.  It was determined that the new rates have at most
a
%EBstart
5\%
%EBend
effect on the ejected abundances of several isotopes, all in cases where a significant portion
of the mass participates in $\alpha$-rich freeze-out.

%---------------------------------------------------------------------------------------------------------------------------------------------
\begin{acknowledgments}
%\label{sec:acknow}

The authors would like to thank the operations staff at the University of Notre Dame for their assistance in the
measurements.  This work was supported by the National Science Foundation under Grant Nos. PHY 1102511,
PHY 08-22648 (Joint Institute for Nuclear Astrophysics) and PHY 0969058. TR was partially supported by the Swiss NSF, the European Research Council, and the THEXO collaboration within the 7$^{th}$ Framework Program ENSAR of the EU.

\end{acknowledgments}

%%%%%%%%%%%%%%%   bibliography   %%%%%%%%%%%%%%%

%


\begin{thebibliography}{25}%
\makeatletter
\providecommand \@ifxundefined [1]{%
 \@ifx{#1\undefined}
}%
\providecommand \@ifnum [1]{%
 \ifnum #1\expandafter \@firstoftwo
 \else \expandafter \@secondoftwo
 \fi
}%
\providecommand \@ifx [1]{%
 \ifx #1\expandafter \@firstoftwo
 \else \expandafter \@secondoftwo
 \fi
}%
\providecommand \natexlab [1]{#1}%
\providecommand \enquote  [1]{``#1''}%
\providecommand \bibnamefont  [1]{#1}%
\providecommand \bibfnamefont [1]{#1}%
\providecommand \citenamefont [1]{#1}%
\providecommand \href@noop [0]{\@secondoftwo}%
\providecommand \href [0]{\begingroup \@sanitize@url \@href}%
\providecommand \@href[1]{\@@startlink{#1}\@@href}%
\providecommand \@@href[1]{\endgroup#1\@@endlink}%
\providecommand \@sanitize@url [0]{\catcode `\\12\catcode `\$12\catcode
  `\&12\catcode `\#12\catcode `\^12\catcode `\_12\catcode `\%12\relax}%
\providecommand \@@startlink[1]{}%
\providecommand \@@endlink[0]{}%
\providecommand \url  [0]{\begingroup\@sanitize@url \@url }%
\providecommand \@url [1]{\endgroup\@href {#1}{\urlprefix }}%
\providecommand \urlprefix  [0]{URL }%
\providecommand \Eprint [0]{\href }%
\providecommand \doibase [0]{http://dx.doi.org/}%
\providecommand \selectlanguage [0]{\@gobble}%
\providecommand \bibinfo  [0]{\@secondoftwo}%
\providecommand \bibfield  [0]{\@secondoftwo}%
\providecommand \translation [1]{[#1]}%
\providecommand \BibitemOpen [0]{}%
\providecommand \bibitemStop [0]{}%
\providecommand \bibitemNoStop [0]{.\EOS\space}%
\providecommand \EOS [0]{\spacefactor3000\relax}%
\providecommand \BibitemShut  [1]{\csname bibitem#1\endcsname}%
\let\auto@bib@innerbib\@empty
%</preamble>
\bibitem [{\citenamefont {K\"{a}ppeler}\ \emph {et~al.}(2011)\citenamefont
  {K\"{a}ppeler}, \citenamefont {Gallino}, \citenamefont {Bisterzo},\ and\
  \citenamefont {Aoki}}]{Kappeler11}%
  \BibitemOpen
  \bibfield  {author} {\bibinfo {author} {\bibfnamefont {F.}~\bibnamefont
  {K\"{a}ppeler}}, \bibinfo {author} {\bibfnamefont {R.}~\bibnamefont
  {Gallino}}, \bibinfo {author} {\bibfnamefont {S.}~\bibnamefont {Bisterzo}}, \
  and\ \bibinfo {author} {\bibfnamefont {W.}~\bibnamefont {Aoki}},\ }\href@noop
  {} {\bibfield  {journal} {\bibinfo  {journal} {Rev. Mod. Phys.}\ }\textbf
  {\bibinfo {volume} {83}},\ \bibinfo {pages} {157} (\bibinfo {year}
  {2011})}\BibitemShut {NoStop}%
\bibitem [{\citenamefont {Arnould}\ \emph {et~al.}(2007)\citenamefont
  {Arnould}, \citenamefont {Goriely},\ and\ \citenamefont
  {Takahashi}}]{Arnould07}%
  \BibitemOpen
  \bibfield  {author} {\bibinfo {author} {\bibfnamefont {M.}~\bibnamefont
  {Arnould}}, \bibinfo {author} {\bibfnamefont {S.}~\bibnamefont {Goriely}}, \
  and\ \bibinfo {author} {\bibfnamefont {K.}~\bibnamefont {Takahashi}},\
  }\href@noop {} {\bibfield  {journal} {\bibinfo  {journal} {Phys. Rep.}\
  }\textbf {\bibinfo {volume} {450}},\ \bibinfo {pages} {97 } (\bibinfo {year}
  {2007})}\BibitemShut {NoStop}%
\bibitem [{\citenamefont {Rauscher}\ \emph {et~al.}(2013)\citenamefont
  {Rauscher}, \citenamefont {Dauphas}, \citenamefont {Dillmann}, \citenamefont
  {Fr\"{o}hlich}, \citenamefont {F\"{u}l\"{o}p},\ and\ \citenamefont
  {Gy.Gy\"{u}rky}}]{Rauscher13}%
  \BibitemOpen
  \bibfield  {author} {\bibinfo {author} {\bibfnamefont {T.}~\bibnamefont
  {Rauscher}}, \bibinfo {author} {\bibfnamefont {N.}~\bibnamefont {Dauphas}},
  \bibinfo {author} {\bibfnamefont {I.}~\bibnamefont {Dillmann}}, \bibinfo
  {author} {\bibfnamefont {C.}~\bibnamefont {Fr\"{o}hlich}}, \bibinfo {author}
  {\bibfnamefont {Z.}~\bibnamefont {F\"{u}l\"{o}p}}, \ and\ \bibinfo {author}
  {\bibnamefont {Gy.Gy\"{u}rky}},\ }\href@noop {} {\bibfield  {journal}
  {\bibinfo  {journal} {Rep. Prog. Phys.}\ }\textbf {\bibinfo {volume} {76}},\
  \bibinfo {pages} {066201} (\bibinfo {year} {2013})}\BibitemShut {NoStop}%
\bibitem [{\citenamefont {The}\ \emph {et~al.}(1998)\citenamefont {The},
  \citenamefont {Clayton}, \citenamefont {Jin},\ and\ \citenamefont
  {Meyer}}]{The98}%
  \BibitemOpen
  \bibfield  {author} {\bibinfo {author} {\bibfnamefont {L.-S.}\ \bibnamefont
  {The}}, \bibinfo {author} {\bibfnamefont {D.}~\bibnamefont {Clayton}},
  \bibinfo {author} {\bibfnamefont {L.}~\bibnamefont {Jin}}, \ and\ \bibinfo
  {author} {\bibfnamefont {B.}~\bibnamefont {Meyer}},\ }\href@noop {}
  {\bibfield  {journal} {\bibinfo  {journal} {Astrophys. J.}\ }\textbf
  {\bibinfo {volume} {504}},\ \bibinfo {pages} {500} (\bibinfo {year}
  {1998})}\BibitemShut {NoStop}%
\bibitem [{\citenamefont {Iliadis}\ \emph {et~al.}(2002)\citenamefont
  {Iliadis}, \citenamefont {Champagne}, \citenamefont {Jos\'e}, \citenamefont
  {Starrfield},\ and\ \citenamefont {Tupper}}]{Iliadis02}%
  \BibitemOpen
  \bibfield  {author} {\bibinfo {author} {\bibfnamefont {C.}~\bibnamefont
  {Iliadis}}, \bibinfo {author} {\bibfnamefont {A.}~\bibnamefont {Champagne}},
  \bibinfo {author} {\bibfnamefont {J.}~\bibnamefont {Jos\'e}}, \bibinfo
  {author} {\bibfnamefont {S.}~\bibnamefont {Starrfield}}, \ and\ \bibinfo
  {author} {\bibfnamefont {P.}~\bibnamefont {Tupper}},\ }\href@noop {}
  {\bibfield  {journal} {\bibinfo  {journal} {Astrophys. J. Suppl.}\ }\textbf
  {\bibinfo {volume} {142}},\ \bibinfo {pages} {105} (\bibinfo {year}
  {2002})}\BibitemShut {NoStop}%
\bibitem [{\citenamefont {Hix}\ \emph {et~al.}(2003)\citenamefont {Hix},
  \citenamefont {Smith}, \citenamefont {Starrfield}, \citenamefont
  {Mezzacappa},\ and\ \citenamefont {Smith}}]{Hix03}%
  \BibitemOpen
  \bibfield  {author} {\bibinfo {author} {\bibfnamefont {W.}~\bibnamefont
  {Hix}}, \bibinfo {author} {\bibfnamefont {M.}~\bibnamefont {Smith}}, \bibinfo
  {author} {\bibfnamefont {S.}~\bibnamefont {Starrfield}}, \bibinfo {author}
  {\bibfnamefont {A.}~\bibnamefont {Mezzacappa}}, \ and\ \bibinfo {author}
  {\bibfnamefont {D.~L.}\ \bibnamefont {Smith}},\ }\href@noop {} {\bibfield
  {journal} {\bibinfo  {journal} {Nucl. Phys. A}\ }\textbf {\bibinfo {volume}
  {718}},\ \bibinfo {pages} {620} (\bibinfo {year} {2003})}\BibitemShut
  {NoStop}%
\bibitem [{\citenamefont {Rapp}\ \emph {et~al.}(2006)\citenamefont {Rapp},
  \citenamefont {G\"{o}rres}, \citenamefont {Wiescher}, \citenamefont
  {Schatz},\ and\ \citenamefont {K\"{a}ppeler}}]{Rapp06}%
  \BibitemOpen
  \bibfield  {author} {\bibinfo {author} {\bibfnamefont {W.}~\bibnamefont
  {Rapp}}, \bibinfo {author} {\bibfnamefont {J.}~\bibnamefont {G\"{o}rres}},
  \bibinfo {author} {\bibfnamefont {M.}~\bibnamefont {Wiescher}}, \bibinfo
  {author} {\bibfnamefont {H.}~\bibnamefont {Schatz}}, \ and\ \bibinfo {author}
  {\bibfnamefont {F.}~\bibnamefont {K\"{a}ppeler}},\ }\href@noop {} {\bibfield
  {journal} {\bibinfo  {journal} {Astrophys. J.}\ }\textbf {\bibinfo {volume}
  {653}},\ \bibinfo {pages} {474} (\bibinfo {year} {2006})}\BibitemShut
  {NoStop}%
\bibitem [{\citenamefont {Bravo}\ and\ \citenamefont
  {Mart\'inez-Pinedo}(2012)}]{Bravo12}%
  \BibitemOpen
  \bibfield  {author} {\bibinfo {author} {\bibfnamefont {E.}~\bibnamefont
  {Bravo}}\ and\ \bibinfo {author} {\bibfnamefont {G.}~\bibnamefont
  {Mart\'inez-Pinedo}},\ }\href@noop {} {\bibfield  {journal} {\bibinfo
  {journal} {Phys. Rev. C}\ }\textbf {\bibinfo {volume} {85}},\ \bibinfo
  {pages} {055805} (\bibinfo {year} {2012})}\BibitemShut {NoStop}%
\bibitem [{\citenamefont {Morinaga}(1956)}]{Morinaga56}%
  \BibitemOpen
  \bibfield  {author} {\bibinfo {author} {\bibfnamefont {H.}~\bibnamefont
  {Morinaga}},\ }\href@noop {} {\bibfield  {journal} {\bibinfo  {journal}
  {Phys. Rev.}\ }\textbf {\bibinfo {volume} {101}},\ \bibinfo {pages} {100}
  (\bibinfo {year} {1956})}\BibitemShut {NoStop}%
\bibitem [{\citenamefont {Ball}\ \emph {et~al.}(1959)\citenamefont {Ball},
  \citenamefont {Fairhall},\ and\ \citenamefont {Halpern}}]{Ball59}%
  \BibitemOpen
  \bibfield  {author} {\bibinfo {author} {\bibfnamefont {J.}~\bibnamefont
  {Ball}}, \bibinfo {author} {\bibfnamefont {A.}~\bibnamefont {Fairhall}}, \
  and\ \bibinfo {author} {\bibfnamefont {I.}~\bibnamefont {Halpern}},\
  }\href@noop {} {\bibfield  {journal} {\bibinfo  {journal} {Phys. Rev.}\
  }\textbf {\bibinfo {volume} {114}},\ \bibinfo {pages} {305} (\bibinfo {year}
  {1959})}\BibitemShut {NoStop}%
\bibitem [{\citenamefont {McGowan}\ \emph {et~al.}(1964)\citenamefont
  {McGowan}, \citenamefont {Stelson},\ and\ \citenamefont {Smith}}]{McGowan64}%
  \BibitemOpen
  \bibfield  {author} {\bibinfo {author} {\bibfnamefont {F.}~\bibnamefont
  {McGowan}}, \bibinfo {author} {\bibfnamefont {P.}~\bibnamefont {Stelson}}, \
  and\ \bibinfo {author} {\bibfnamefont {W.}~\bibnamefont {Smith}},\
  }\href@noop {} {\bibfield  {journal} {\bibinfo  {journal} {Phys. Rev.}\
  }\textbf {\bibinfo {volume} {133}},\ \bibinfo {pages} {907} (\bibinfo {year}
  {1964})}\BibitemShut {NoStop}%
\bibitem [{HIB()}]{HIBAL}%
  \BibitemOpen
  \href@noop {} {}\bibinfo {note} {Hope College Ion Beam Analysis Laboratory,
  www.hope.edu/academic/physics/facilities/accelerator/ (2014)}\BibitemShut
  {NoStop}%
\bibitem [{\citenamefont {Mayer}()}]{Mayer97}%
  \BibitemOpen
  \bibfield  {author} {\bibinfo {author} {\bibfnamefont {M.}~\bibnamefont
  {Mayer}},\ }\href@noop {} {}\bibinfo {note} {\textit{SIMNRA User's Guide},
  Report IPP 9/113, Max-Planck-Institut f\"{u}r Plasmaphysik, Garching,
  Germany, 1997}\BibitemShut {NoStop}%
\bibitem [{\citenamefont {Tarasov}\ and\ \citenamefont
  {Bazin}(2008)}]{LISE}%
  \BibitemOpen
  \bibfield  {author} {\bibinfo {author} {\bibfnamefont {O.B.}~\bibnamefont
  {Tarasov}}\ and\ \bibinfo {author} {\bibfnamefont {D.}\ \bibnamefont
  {Bazin}},\ }\href@noop {} {\bibfield  {journal} {\bibinfo  {journal}
  {Nucl. Instr. Meth. B}\ }\textbf {\bibinfo {volume} {266}},\
  \bibinfo {pages} {4657} (\bibinfo {year} {2008})}\BibitemShut {NoStop}%
\bibitem [{\citenamefont {Simon}\ \emph {et~al.}(2013)\citenamefont {Simon},
  \citenamefont {Quinn}, \citenamefont {Spyrou}, \citenamefont {Battaglia},
  \citenamefont {Beskin}, \citenamefont {Best}, \citenamefont {Bucher},
  \citenamefont {Couder}, \citenamefont {DeYoung}, \citenamefont {Fang},
  \citenamefont {Görres}, \citenamefont {Kontos}, \citenamefont {Li},
  \citenamefont {Liddick}, \citenamefont {Long}, \citenamefont {Lyons},
  \citenamefont {Padmanabhan}, \citenamefont {Peace}, \citenamefont {Roberts},
  \citenamefont {Robertson}, \citenamefont {Smith}, \citenamefont {Smith},
  \citenamefont {Stech}, \citenamefont {Stefanek}, \citenamefont {Tan},
  \citenamefont {Tang},\ and\ \citenamefont {Wiescher}}]{Simon13}%
  \BibitemOpen
  \bibfield  {author} {\bibinfo {author} {\bibfnamefont {A.}~\bibnamefont
  {Simon}}, \bibinfo {author} {\bibfnamefont {S.~J.}\ \bibnamefont {Quinn}},
  \bibinfo {author} {\bibfnamefont {A.}~\bibnamefont {Spyrou}}, \bibinfo
  {author} {\bibfnamefont {A.}~\bibnamefont {Battaglia}}, \bibinfo {author}
  {\bibfnamefont {I.}~\bibnamefont {Beskin}}, \bibinfo {author} {\bibfnamefont
  {A.}~\bibnamefont {Best}}, \bibinfo {author} {\bibfnamefont {B.}~\bibnamefont
  {Bucher}}, \bibinfo {author} {\bibfnamefont {M.}~\bibnamefont {Couder}},
  \bibinfo {author} {\bibfnamefont {P.}~\bibnamefont {DeYoung}}, \bibinfo
  {author} {\bibfnamefont {X.}~\bibnamefont {Fang}}, \bibinfo {author}
  {\bibfnamefont {J.}~\bibnamefont {Görres}}, \bibinfo {author} {\bibfnamefont
  {A.}~\bibnamefont {Kontos}}, \bibinfo {author} {\bibfnamefont
  {Q.}~\bibnamefont {Li}}, \bibinfo {author} {\bibfnamefont {S.}~\bibnamefont
  {Liddick}}, \bibinfo {author} {\bibfnamefont {A.}~\bibnamefont {Long}},
  \bibinfo {author} {\bibfnamefont {S.}~\bibnamefont {Lyons}}, \bibinfo
  {author} {\bibfnamefont {K.}~\bibnamefont {Padmanabhan}}, \bibinfo {author}
  {\bibfnamefont {J.}~\bibnamefont {Peace}}, \bibinfo {author} {\bibfnamefont
  {A.}~\bibnamefont {Roberts}}, \bibinfo {author} {\bibfnamefont
  {D.}~\bibnamefont {Robertson}}, \bibinfo {author} {\bibfnamefont
  {K.}~\bibnamefont {Smith}}, \bibinfo {author} {\bibfnamefont
  {M.}~\bibnamefont {Smith}}, \bibinfo {author} {\bibfnamefont
  {E.}~\bibnamefont {Stech}}, \bibinfo {author} {\bibfnamefont
  {B.}~\bibnamefont {Stefanek}}, \bibinfo {author} {\bibfnamefont
  {W.}~\bibnamefont {Tan}}, \bibinfo {author} {\bibfnamefont {X.}~\bibnamefont
  {Tang}}, \ and\ \bibinfo {author} {\bibfnamefont {M.}~\bibnamefont
  {Wiescher}},\ }\href@noop {} {\bibfield  {journal} {\bibinfo  {journal}
  {Nucl. Instr. Meth. A}\ }\textbf {\bibinfo {volume} {703}},\ \bibinfo {pages}
  {16} (\bibinfo {year} {2013})}\BibitemShut {NoStop}%
\bibitem [{\citenamefont {Prokop}\ and\ \citenamefont
  {Liddick}(2014)}]{Prokop14}%
  \BibitemOpen
  \bibfield  {author} {\bibinfo {author} {\bibfnamefont {C.}~\bibnamefont
  {Prokop}}\ and\ \bibinfo {author} {\bibfnamefont {S.}~\bibnamefont
  {Liddick}},\ }\href@noop {} {\bibfield  {journal} {\bibinfo  {journal} {Nucl.
  Instr. Meth. A}\ }\textbf {\bibinfo {volume} {741}},\ \bibinfo {pages} {163}
  (\bibinfo {year} {2014})}\BibitemShut {NoStop}%
\bibitem [{\citenamefont {Spyrou}\ \emph {et~al.}(2007)\citenamefont {Spyrou},
  \citenamefont {Becker}, \citenamefont {Lagoyannis}, \citenamefont
  {Harissopulos},\ and\ \citenamefont {Rolfs}}]{Spyrou07}%
  \BibitemOpen
  \bibfield  {author} {\bibinfo {author} {\bibfnamefont {A.}~\bibnamefont
  {Spyrou}}, \bibinfo {author} {\bibfnamefont {H.~W.}\ \bibnamefont {Becker}},
  \bibinfo {author} {\bibfnamefont {A.}~\bibnamefont {Lagoyannis}}, \bibinfo
  {author} {\bibfnamefont {S.}~\bibnamefont {Harissopulos}}, \ and\ \bibinfo
  {author} {\bibfnamefont {C.}~\bibnamefont {Rolfs}},\ }\href@noop {}
  {\bibfield  {journal} {\bibinfo  {journal} {Phys. Rev. C}\ }\textbf {\bibinfo
  {volume} {76}},\ \bibinfo {pages} {015802} (\bibinfo {year}
  {2007})}\BibitemShut {NoStop}%
\bibitem [{\citenamefont {Vlieks}\ \emph {et~al.}(1974)\citenamefont {Vlieks},
  \citenamefont {Morgan},\ and\ \citenamefont {Blatt}}]{Vlieks74}%
  \BibitemOpen
  \bibfield  {author} {\bibinfo {author} {\bibfnamefont {A.}~\bibnamefont
  {Vlieks}}, \bibinfo {author} {\bibfnamefont {J.}~\bibnamefont {Morgan}}, \
  and\ \bibinfo {author} {\bibfnamefont {S.}~\bibnamefont {Blatt}},\
  }\href@noop {} {\bibfield  {journal} {\bibinfo  {journal} {Nucl. Phys. A}\
  }\textbf {\bibinfo {volume} {224}},\ \bibinfo {pages} {492 } (\bibinfo {year}
  {1974})}\BibitemShut {NoStop}%
\bibitem [{\citenamefont {Rauscher}()}]{SMARAGD}%
  \BibitemOpen
  \bibfield  {author} {\bibinfo {author} {\bibfnamefont {T.}~\bibnamefont
  {Rauscher}},\ }\href@noop {} {}\bibinfo {note} {Code \textit{SMARAGD},
  version 0.9.0s (2012)}\BibitemShut {NoStop}%
\bibitem [{\citenamefont {Rauscher}\ and\ \citenamefont
  {Thielemann}\ and\ \citenamefont {Kratz}(1997)}]{prc56}%
  \BibitemOpen
  \bibfield  {author} {\bibinfo {author} {\bibfnamefont {T.}~\bibnamefont
  {Rauscher}}\ and\ \bibinfo {author} {\bibfnamefont {F.-K.}\ \bibnamefont
  {Thielemann}}\ and\ \bibinfo {author} {\bibfnamefont {K.-L.}\ \bibnamefont
  {Kratz}},\ }\href@noop {} {\bibfield  {journal} {\bibinfo  {journal}
  {Phys. Rev. C}\ }\textbf {\bibinfo {volume} {56}},\
  \bibinfo {pages} {1613} (\bibinfo {year} {1997})}\BibitemShut {NoStop}%
\bibitem [{\citenamefont {Rauscher}(2010)}]{energywindows}%
  \BibitemOpen
  \bibfield  {author} {\bibinfo {author} {\bibfnamefont {T.}~\bibnamefont
  {Rauscher}},\ }\href@noop {} {\bibfield  {journal} {\bibinfo  {journal}
  {Phys. Rev. C}\ }\textbf {\bibinfo {volume} {81}},\ \bibinfo {pages} {045807}
  (\bibinfo {year} {2010})}\BibitemShut {NoStop}%
\bibitem [{\citenamefont {Rauscher}(2012)}]{Rauscher12}%
  \BibitemOpen
  \bibfield  {author} {\bibinfo {author} {\bibfnamefont {T.}~\bibnamefont
  {Rauscher}},\ }\href@noop {} {\bibfield  {journal} {\bibinfo  {journal}
  {Astrophys. J. Suppl.}\ }\textbf {\bibinfo {volume} {201}},\ \bibinfo {pages}
  {26} (\bibinfo {year} {2012})}\BibitemShut {NoStop}%
\bibitem [{\citenamefont {Hauser}\ and\ \citenamefont
  {Feshbach}(1952)}]{haufesh}%
  \BibitemOpen
  \bibfield  {author} {\bibinfo {author} {\bibfnamefont {W.}~\bibnamefont
  {Hauser}}\ and\ \bibinfo {author} {\bibfnamefont {H.}~\bibnamefont
  {Feshbach}},\ }\href@noop {} {\bibfield  {journal} {\bibinfo  {journal}
  {Phys. Rev.}\ }\textbf {\bibinfo {volume} {87}},\ \bibinfo {pages} {366}
  (\bibinfo {year} {1952})}\BibitemShut {NoStop}%
\bibitem [{\citenamefont {Rauscher}(2011)}]{raureview}%
  \BibitemOpen
  \bibfield  {author} {\bibinfo {author} {\bibfnamefont {T.}~\bibnamefont
  {Rauscher}},\ }\href@noop {} {\bibfield  {journal} {\bibinfo  {journal} {Int.
  J. Mod. Phys. E}\ }\textbf {\bibinfo {volume} {20}},\ \bibinfo {pages} {1071}
  (\bibinfo {year} {2011})}\BibitemShut {NoStop}%
\bibitem [{\citenamefont {McFadden}\ and\ \citenamefont
  {Satchler}(1966)}]{McFadden66}%
  \BibitemOpen
  \bibfield  {author} {\bibinfo {author} {\bibfnamefont {L.}~\bibnamefont
  {McFadden}}\ and\ \bibinfo {author} {\bibfnamefont {G.}~\bibnamefont
  {Satchler}},\ }\href@noop {} {\bibfield  {journal} {\bibinfo  {journal}
  {Nucl. Phys.}\ }\textbf {\bibinfo {volume} {84}},\ \bibinfo {pages} {177}
  (\bibinfo {year} {1966})}\BibitemShut {NoStop}%
\bibitem [{\citenamefont {Rauscher}\ and\ \citenamefont
  {Thielemann}(2000)}]{Rauscher00}%
  \BibitemOpen
  \bibfield  {author} {\bibinfo {author} {\bibfnamefont {T.}~\bibnamefont
  {Rauscher}}\ and\ \bibinfo {author} {\bibfnamefont {F.-K.}\ \bibnamefont
  {Thielemann}},\ }\href@noop {} {\bibfield  {journal} {\bibinfo  {journal}
  {At. Data and Nucl. Data Tables}\ }\textbf {\bibinfo {volume} {75}},\
  \bibinfo {pages} {1} (\bibinfo {year} {2000})}\BibitemShut {NoStop}%
\bibitem [{\citenamefont {Cyburt}\ \emph {et~al.}(2010)\citenamefont {Cyburt},
  \citenamefont {Amthor}, \citenamefont {Ferguson}, \citenamefont {Meisel},
  \citenamefont {Smith}, \citenamefont {Warren}, \citenamefont {Heger},
  \citenamefont {Hoffman}, \citenamefont {Rauscher}, \citenamefont {Sakharuk},
  \citenamefont {Schatz}, \citenamefont {Thielemann},\ and\ \citenamefont
  {Wiescher}}]{Cyburt10}%
  \BibitemOpen
  \bibfield  {author} {\bibinfo {author} {\bibfnamefont {R.~H.}\ \bibnamefont
  {Cyburt}}, \bibinfo {author} {\bibfnamefont {A.~M.}\ \bibnamefont {Amthor}},
  \bibinfo {author} {\bibfnamefont {R.}~\bibnamefont {Ferguson}}, \bibinfo
  {author} {\bibfnamefont {Z.}~\bibnamefont {Meisel}}, \bibinfo {author}
  {\bibfnamefont {K.}~\bibnamefont {Smith}}, \bibinfo {author} {\bibfnamefont
  {S.}~\bibnamefont {Warren}}, \bibinfo {author} {\bibfnamefont
  {A.}~\bibnamefont {Heger}}, \bibinfo {author} {\bibfnamefont {R.~D.}\
  \bibnamefont {Hoffman}}, \bibinfo {author} {\bibfnamefont {T.}~\bibnamefont
  {Rauscher}}, \bibinfo {author} {\bibfnamefont {A.}~\bibnamefont {Sakharuk}},
  \bibinfo {author} {\bibfnamefont {H.}~\bibnamefont {Schatz}}, \bibinfo
  {author} {\bibfnamefont {F.~K.}\ \bibnamefont {Thielemann}}, \ and\ \bibinfo
  {author} {\bibfnamefont {M.}~\bibnamefont {Wiescher}},\ }\href@noop {}
  {\bibfield  {journal} {\bibinfo  {journal} {Astrophys. J. Suppl.}\ }\textbf
  {\bibinfo {volume} {189}},\ \bibinfo {pages} {240} (\bibinfo {year}
  {2010})}\BibitemShut {NoStop}%
\end{thebibliography}
\end{document}